\documentclass[twocolumn,prl,showpacs]{revtex4}

\usepackage{bm}
\usepackage{amsmath}
\usepackage{graphicx}

\begin{document}

%
%
\title{Optical Multi-hysteresises and "Rogue Waves" in Nonlinear Plasma}
\author{A.~E.\ Kaplan}
\affiliation{Dept. of Electrical and Computer Engineering, The Johns Hopkins University, Baltimore, MD 21218}

\date{December 30, 2009}

\begin{abstract}
An overdense plasma layer irradiated by an intense light
can exhibit dramatic nonlinear-optical effects
due to a relativistic mass-effect of free electrons: 
highly-multiple hysteresises of reflection and transition, 
and emergence of gigantic "rogue waves". Those are 
trapped quasi-soliton field spikes inside the 
layer, sustained by an incident radiation with a tiny 
fraction of their peak intensity once they have been 
excited by orders of magnitude larger pumping.  The 
phenomenon persists even in the layers with "soft" boundaries, 
as well as in a semi-infinite plasma with low absorption.
\end{abstract}


\preprint{Submitted to \emph{Phys.~Rev.~Lett.}}

\maketitle


Diverse wave-related phenomena
exhibit a common critical behavior:
a dramatic transition at a crossover point
from  a traveling wave in the underlying
(semi-infinite)  medium to a non-propagating,
evanescent wave that carry no energy.
The crossover occurs in optics
at the angle of total internal reflection at 
a dielectric interface,
at a laser frequency near either a plasma frequency,
or a waveguide cut-off frequency, or band-gap edge of
a material, including BEC;
in quantum mechanics for electron scattering at 
the energy close to a potential plateau, etc.
It can be of great significance
to \emph{nonlinear} optics:
a nonlinear refractive index
can cause a phase-transition-like effect,
since a small light-induced change
may translate into a switch from full reflection to full
transmission, resulting in a huge hysteresis.
Predicted in [1] for nonlinear interfaces,
it was explored experimentally in [2],
with an inconclusive outcome,
with some of the experiments (including the latest [2c])
showing a clear hysteresis, 
while [2b] showing none (see also below).
2D numerical simulations were 
not well suited then for modeling hysteresis;
their very formulation excluded multivalued outcome
by using single-valued boundary conditions.

In this Letter we discover, however, that 
even the most basic, 1D-case, reveals a large and 
apparently little known phenomenon
of highly-multistable nonlinear EM-propagation
and the emergence of trapped "rogue" (R) waves,
with intensity exceeding the incident one by orders of magnitude.
They may be released by plasma expansion,
in a phase-like transition, as in a boiling liquid.
Even a slight nonlinearity due to 
the most fundamental mechanism
-- relativistic (RL) mass-effect of free electrons -- 
suffices to initiate the effect.
Multiple (up to hundreds) hysteretic
jumps between almost full reflection and full transparency
may occur as the laser intensity is swept up and down.
We treat the problem here in the context of ideal
plasma on account of recent interest
in high-intensity laser-plasma interactions
and fundamental nature of RL-nonlinearity, 
but all approaches are valid for other crossover problems.
Temporal RL-solitons have been
considered in detail in the literature [3];
close to those of underdense plasma
are the so called Bragg or band-gap
solitions [4], including the ones in BEC [5].
The difference in this work is made by 
multi-hysteresises (and \emph{standing, unmoving}
quasi-solitons instead of propagating ones)
due to self-induced retro-reflection.
The remarkable
new property is that for the \emph{same incident power}
an EM-wave can penetrate into a nonlinear material to
\emph{different depths} -- varying by orders of magnitude --
depending of the history of pumping.
We assume a stationary, $cw$, or long pulse mode, 
and use only RL-nonlinearity in a cold plasma. 
While this model is greatly simplified \emph{vs}
various kinetic approaches, it allows us to keep the 
basic features necessary to elucidate new results,
and have the theory applicable to other systems.

Even a few-$\lambda$-thick plasma layer can produce the effect, 
so that the absorption 
of light (including nonlinear self-focusing)
would not affect the propagation significantly.
Furthermore, while abrupt boundaries 
contribute to the effect, they are not essential: 
a layer with "soft" boundaries still exhibits all 
the major features of the phenomenon.
The phenomenon persists even for \emph{semi-infinite}
plasma with a small absorption, which
also develops a strong retro-reflection.
The "Sommerfeld condition" (no wave comes
from the "infinity") is to be revisited  here:
a wave \emph{is back-reflected} 
deep inside the plasma and returns to the boundary.
Imposed on a forward wave, it results in a semi-standing wave,
trapped quasi-soliton R-waves,
and multistability, same as in a finite layer.
The energy accumulated in R-waves and excited free electrons,
can then be released if plasma density reduces,
as it may happen, e. g. in an astrophysical environment.

The wave propagation here
is governed by the same so called nonlinear 
Klein-Gordon-Fock equation:
\begin{equation}
{\bf [} \nabla^2  -  
\partial^2 / ( {\bf v} \partial t )^2 {\bf ]} \psi  = 
k_0^2  f (  | \psi |^2  ) \psi  ;  \ \ \  f ( 0 ) = 1 .
\tag{1}
\end{equation}
where $f (  | \psi |^2   )$ is a function responsible for nonlinearity.
Here, a generic variable, $\psi$, could be a scalar 
(e.g. a wave function in RQM), or a field vector
in EM-wave propagation, ${\bf v}$ is a scale velocity,
(${\bf v = c}$ for plasma and in RQM),
and $k_0^{-1}$ is a spatial scale of the problem,
(for RQM, $k_0 = m_0 c / \hbar = 2 \pi / \lambda_C$,
where $\lambda_C$ is the Compton
wavelength, and $m_0$ is the rest-mass
of an electron); 
for plasma $k_0 = \omega_{pl} / c$,
where $\omega_{pl}$ is a plasma frequency
due to free-charge density, $\rho_e$,
and in the X-ray physics,
$\omega_{ph} = E_{ph} / \hbar$ is related to 
photo-ionization limit, $E_{ph}$, of atoms.
For an $\omega$-monochromatic wave 
$\psi = \mathcal{\vec{E}} ( \vec{r} )$ $e^{ - i \omega t } / 2 + c. c.$,
where $\mathcal{\vec{E}} ( \vec{r} )$ is 
a complex amplitude,
and using $u^2 = | \mathcal{E} |^2$,
(1) is reduced to a nonlinear Helmholtz equation:
\begin{equation}
\nabla^2 \mathcal{\vec{E}} +  k^2 \epsilon ( u^2 )  \mathcal{\vec{E}} = 0 ;
\ \ \
\epsilon_{pl}  = 1  - \omega_{pl}^2 ( u^2 ) / \omega^2
\tag{2}
\end{equation}
where $k = \omega / c$, and
$\omega_{pl}^2  = 4 \pi e \rho_e / m $,
$m$ -- electron mass,
and $\mathcal{E} = E / E_{nl}$ 
with $E_{nl}$ being some characteristic nonlinear scale;
for a RL-mass-effect it is $E_{rl} = \omega m_0 c / e$.

In general, nonlinear $\epsilon$ may
have various origins: a varying ionization rate,
plasma waves, ponderomotive force, etc. 
Assuming fully ionized gas, $\rho_e$ $=$ $const$,
and a circularly-polarized wave,
$\mathcal{E} ( \zeta ) $ $( \hat{e}_x  + i
\hat{e}_y )$ $ e^{{-i} \omega t} /2  + c.c.$,
that has very negligible high-harmonics 
generation and minimal longitudinal plasma waves excitation,
the most basic remaining source of nonlinearity 
is a field-induced RL mass-effect of electron:
$m = m_0 \gamma$, with a relativistic factor
$\gamma =  \sqrt { 1 + ( p / m_0 c )^2 } =$ $\sqrt
{ 1 + u^2}$ [see e. g. [6])
where $p$ is the momentum of electron, so that
\begin{equation}
\epsilon_{rl}  = 1  -  [ \nu^2 \gamma ( u^2 ) ]^{-1}
   \qquad\mbox{with}\quad     \nu  =  \omega / ( \omega_{pl} )_0
\tag{3}
\end{equation}
where $( \omega_{pl} )_0$ 
is a linear plasma frequency with $m = m_0$.
Since $m = m ( u^2 )$, a single electron 
exhibits large hysteretic cyclotron resonance 
predicted in [6a] and observed in [7].
The mass-effect has also became one of the major players
of light-plasma interaction [3],
e. g. in RL self-focusing,
and in acceleration of electrons by the 
beat-wave and wake-field.
The EM-propagation could also
be accompanied by RL-intrinsic bistability [8].

In a 1D-case, letting a plane EM-wave propagate
in the $z$-axis, we have $\nabla^2 = d^2 / d z^2$.
For a boundary between two dielectrics, a EM-wave 
incident from a dielectric with $\epsilon_{in}  > 0$
under the angle $\theta$ 
onto a material of $\epsilon_{NL} > 0$,
$\epsilon$ in (2) is replaced by 
$ \epsilon_{in} [ \epsilon_{NL} ( \omega ) 
/ \epsilon_{in} ( \omega ) -$ $ \sin^2 \theta ]$.
For a $mw$ waveguide 
with a critical frequency $\omega_{wg}$,
$\epsilon$ in (2) is replaced by 
$\epsilon_{wg} ( 1 - \omega_{wg}^2 / \omega^2 )$.
The crossover point is attained at $\epsilon = 0$.
In this approximation, (2) reduces to
\begin{equation}
\mathcal{E} ^{\prime \prime} + 
\epsilon  ( \zeta ,  u^2 )  \mathcal{E}  = 0 ,
\tag{4}
\end{equation}
where $\zeta = k z $, and "prime" denotes $d / d \zeta$;
in general, we do not assume $\epsilon$ 
$uniform$ in $\zeta$-axis.
In a weakly-nonlinear media
one can break the field into 
counter-propagating traveling waves
and find their amplitudes $via$ boundary conditions.
However, near a crossover point
one in general cannot distinguish between those waves.
To make no assumptions about the wave composition,
we represent the field using real variables
$u$, and phase (eikonal), $\phi$, as
$\mathcal{E} = u ( \zeta ) \it\hbox{exp} {\bf [} i \phi ( \zeta ) {\bf ]}$.
Since $\mathcal{E}$ is in general complex,
while $\epsilon = \epsilon ( u^2 )$,
Eq. (4) is isomorphous to a 3-rd order
eqn for $u$; yet, it is fully integrable in quadratures.
Its first integral is 
a scaled momentum flux of EM-field, 
$P \equiv u^2  \phi ^\prime  = inv$.
In a lossless media $P$ is conserved
over the entire space $ \zeta  < \infty$,
even if the medium is non-uniform,
multi-layered, linear and/or nonlinear, etc.
If a layer borders a dielectric of $\epsilon = \epsilon_{ex}$
at the exit, we have 
$P = u_{ex}^2 \sqrt {\epsilon_{ex}}$,
where $u_{ex}^2$  is the exit wave intensity.
Eq. (2) is reduced then
to a 2-nd order equation for $u$: 
\begin{equation}
u ^{\prime \prime} +  u   {\bf [} \epsilon ( \zeta , u^2 ) -  
P^2 / u^4 {\bf ]}  = 0 ,
\tag{5}
\end{equation}
which makes an unusual yet greatly useful tool.
By addressing only a real amplitude, while 
using flux $P$ as a parameter,
(5) is nonlinear even for a \emph{ linear propagation},
yet is still analytically solvable
if a density $\rho_e$ 
is uniform across the layer 
($ \partial \epsilon / \partial \zeta = 0$).
A full-energy-like invariant of (5) is
${{u ^\prime}^2  / 2} + U ( u^2 )\! = W\! =\! inv $, 
with $U  = {\bf [} \int_0^{u^2} \epsilon ( u^2 )$ $
d ( u^2 ) + $ ${P^2 / u^2 } {\bf ]} / 2$ ,     
where ${{u ^\prime}^2 } / 2$ is "kinetic",
and $U $ -- "potential" energies.
For a RL-nonlinearity (3),
$U ( u^2 ) = ( u^2 + P^2 / u^2 ) / 2 - 
{\bf [} \gamma ( u^2 ) - 1 {\bf ]} / \nu^2$.
Here $W$ is a scaled free EM energy density 
of $\epsilon$-nonlinear medium [9]
$ W = c {\bf [} H^2 + \int \epsilon d ( E^2 ) {\bf ]} / ( 2  E_{rl}^2 ) $,
where $H$ is magnetic field.
If a layer exit wall is a dielectric, 
one has $W  =$ $ U ( u_{ex}^2 )$,
since then $u ^\prime_{ex} = 0$ (see below).
For a metallic mirror,
$W  = {{u_{ex} ^\prime}}^2 / 2$,
since now $u_{ex} = 0$; and $W  = 0$ for
an evanescent wave in a semi-infinite medium.
The implicit solution for spatial dynamics
of $u$ in general case is found now as
$\zeta = $ $\int  \{ 2   [ W  $ $- U ( u^2 ) ] \}^{- 1/2} d u$ .

Boundary conditions at the borders with
linear dielectrics at the entrance, $\zeta = 0$,
with $\epsilon_{in}$,
and at the exit, $\zeta = d$, with $\epsilon_{ex}$, result in
complex amplitudes of incident, $\mathcal{E}_{in}$,
and reflected, $\mathcal{E}_{rfl}$, waves at $\zeta = 0$:
\begin{equation}
\mathcal{E}_{{in} , rfl} =  {\bf [} u  \pm  
\epsilon_{in}^{-1/2}  ( P /  u  -  
i u ^\prime ) {\bf ]} / 2 ;
\tag{6}
\end{equation}
where "$+$" corresponds to $\mathcal{E}_{in}$, 
and "$-$" -- to $\mathcal{E}_{rfl}$,
and the exit point, $\zeta = d$:
$u  =  \mathcal{E}_{ex}   \equiv  u_{ex}$; 
$u ^\prime  = 0$; and $\phi ^\prime  = \sqrt  {\epsilon_{ex}}$.
A condition $P = 0$ corresponds to full reflection,
resulting in either strictly standing wave,
or nonlinear evanescent wave in a semi-infinite
plasma, in particular in a "standing" soliton-like
solution (see below).
If  $\epsilon ( u^2 = 0 ) < 0$, 
there are no $linear$ traveling waves;
yet a purely traveling %
\it nonlinear %
\rm wave
may exist at sufficiently strong intensity 
$u^2 =$ $u_{trv}^2 =$ $const$,
such that $\epsilon ( u_{trv}^2 ) > 0$:
\begin{equation}
\mathcal{E}_{trv} = u_{trv}  \exp ( i \phi ^\prime \zeta ) , 
\  \  \      \phi ^\prime = \pm \sqrt {\epsilon ( u_{trv}^2 )} ,
\tag{7}
\end{equation}
propagating either forward ($+$) or backward ($-$).
However, if $\epsilon ( u =  0 ) < 0$, it is strongly unstable.
A non-periodic solution of (5) with $P = 0$ is a 
nonlinear evanescent wave that forms a standing, trapped soliton.
In low-RL case,
one needs a small detuning from crossover point,
$\delta \equiv \nu - 1  \ll 1$,
to attain the effect at low laser intensity, $u^2 \ll 1$, so that 
the the dielectric constant (3) is Kerr-like and small:
$\epsilon_{rl} \approx - 2 \delta + u^2 / 2$, $| \epsilon_{rl} |  \ll 1$.
A full solution of (5)
with $P = 0$ and $u \to 0$ at $ \zeta  \to \infty$
yields then a standing soliton with a familiar intensity profile:
\begin{equation}
u^2 = {8 \delta} / {\ \it\hbox{cosh}^2 {\bf [} ( \zeta  - \zeta_{pk} )
\sqrt {2 \delta } {\bf ]}}
\tag{8}
\end{equation}
where the peak location $\zeta_{pk}$ is an integration constant.
For an arbitrary frequency, $\nu < 1$, 
the soliton peak intensity is
$u_{sol}^2 = 4 ( 1 - \nu^2 ) / \nu^4$,
instead of $8 \delta$ as in (8);
$\epsilon ( u_{pk}^2 ) =$ 
$( 1 - \nu^2) / ( 2  -  \nu^2 ) \ge 0$.
When $\nu^2  < 1/2$,
it is a strongly-RL soliton,
$u_{sol}^2 \gg 1$, and its peak
narrows down to a half-wave:
$u^2 \approx$ $u_{sol}^2$ 
$\cos^2 ( \zeta  - \zeta_{pk} )$
at $u^2 > 1$.

In a finite layer
one has a mix of standing/evanescent
and traveling waves, with 
$u_{min}^2 = u_{ex}^2 = P > 0$.
A full integration of (5) with nonlinearity 
$\epsilon_{rl} \approx - 2 \delta + u^2 / 2$
yields then elliptic integrals of 
imaginary argument of the first kind;
more importantly, eq. (5) and 
its invariant avails themselves to detailed analysis.
The numerical simulations are needed, 
however, to find a solution for
(a) strongly-RL field [using (5) and its integrals],
or (b) non-uniform plasma density in (5),
or(c) plasma with absorption [eq. (4)].
It then is found by an "inverse propagation" procedure,
whereby we essentially back-track the 
propagation from a purely traveling
exit wave back to the entrance.
One sets first a certain magnitude of 
$u_{ex}^2 = P $, $u_{ex}^{\prime} = 0 $ at the exit,
numerically computes an amplitude profile $u ( \zeta )$
back to the entrance and incident and reflected intensities 
$u_{ex}^2$ and $u_{rfl}^2$ using (6), 
and then maps $u_{ex}^2$ and $u_{rfl}^2$
$vs$ incident intensity, $u_{in}^2$.
A data set $u_{in}^2 ( P )$ and $u_{rfl}^2 ( P )$
for any given $P$ is found then with a single run,
$vs$ a so called multi-shooting
commonly used in search of solution
with conditions set at two boundaries.
This provides a very fast numerical 
simulation $vs$ multi-shooting;
besides, the latter one is very unreliable
when dealing with apriory unknown number of multi-solutions.

For a fully-RL simulation with a $L = 10 \lambda$,
where $L$ is the layer thickness,
Fig. 1 show the emergence
of large number, $N_{hs}$, of huge hysteretic loops 
of the transmission (same as in reflection, not shown here),
which bounces between full transparency
(near the points touching an $FT$ line) to
nearly full reflection (near the points touching an envelope $NFR$).
In general, $N_{hs} = O ( L / \lambda )$.
\begin{figure}
\includegraphics[angle=270,width=3in]{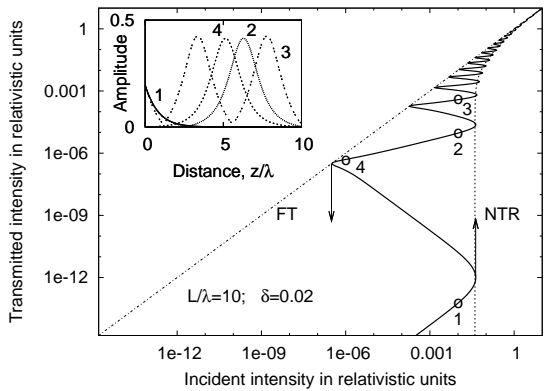}
\caption{Hysteretic transmission of light through a plasma
layer of thickness $L$. $\textbf{FT}$ and $\textbf{NTR}$
are full transparency and near total reflection limits.
Points 1, 2, 3 mark a linear evanescent wave,
1-st, and 2-nd upper stable states respectively, 
and 4 -- a R-wave sustained by a very low pumping.
Arrows indicate direction of jumps within
the lowest hysteretic loop.
Inset: spatial amplitude profiles of waves 
corresponding to points 1-4 in the main plot.}
\end{figure}
In an unbound plasma, the solution
of (5) with a traveling component, $P > 0$,
is a spatially periodic and positively defined,
with the  intensity, $u^2 ( \zeta )$,
bouncing between two limits, $u_{ex}^2$,
and $u_{pk}^2$. If $P / 16  \ll \delta^2  \ll 1$, we have
\begin{equation}
P = u_{ex}^2 \le u^2 \le u_{pk}^2 \approx 
8 \delta  + P / 2 \delta 
\tag{9}
\end{equation}
i. e. the peaks are relatively large,
$u_{pk}^2 \gg u_{ex}^2$ and form
a train of well separated 
quasi-solitons nearly coinciding with a
standing soliton (8) of the peak intensity
$u_{pk}^2 \approx 8 \delta$. 
As $P$ and $u_{in}^2$ increase, they grow 
larger and closer to each other.
The spatial period, $\Lambda$, 
of this structure is:
\begin{equation}
\Lambda / \lambda \approx 
ln ( {16 \delta } / \sqrt P ) / ( 2 \pi \sqrt { 2 \delta }) 
\tag{10}
\end{equation}
In a strongly RL case, $P > 1$,
we have $u_{pk}^2 / P \approx 1 + ( 1 + P )^{-2} $,
and $\Lambda  \approx \lambda / 2$, 
as for a standing, albeit inhibited wave in free space, 
while traveling wave emerges dominant, 
resulting in self-induced transparency.

Hysteretic jumps
occur when either valley or peak of the intensity
profile coincide with an entrance, $\zeta = 0$.
The valley, $u_0^2 =  u_{ex}^2$,
marks an off-jump 
in Fig. 1, and the peak, $u_0^2 =  u_{pk}^2$ -- an on-jump.
Suppose that in a layer of
$L >  \lambda / ( 2 \pi \sqrt { 2 \delta } )$,
the incident intensity $u_{in}^2$ is ramped up from zero. 
When $u_{in}^2 < 2 \delta$, Fig. 1. point 1,
the amplitude is almost exponentially decaying,
$\zeta = 0$, as $u  \approx 2 u_{in} $ 
$ \exp {( - \sqrt {2 \delta} \zeta ) }$, i. e. is
a nearly-linear evanescent wave, curve 1 in Fig. 1 inset; 
the layer is strongly refractive, and the transmission is low.
As $u_{in}^2$ increases,
the front end of that profile swells up,
becoming a semi-bell-like curve, 
close to (8) with $\zeta_{pk} \approx 0$.
With further slight increase
of pumping, it gets unsustainable, and
the field configuration has to jump up to the next stable
branch of excitation, whereby it forms a 
steady R-wave at the back of the layer. 
If after that $u_{in}^2$ is 
pulled down adiabatically slow, the R-wave
moves to the middle of the layer (Fig. 1. point 2,
curve 2 in the inset).
Finally, when it is exactly at the midlayer
(Fig. 1. point 4, curve 4 in the inset),
both valleys are at the borders of the layer,
the pumping is nearly minimal
to support an R-wave; below it
the profile is unsustainable again, and the system
jumps down to a regular nearly-evanescent 
wave and almost full reflection.

At this remarkable point, the layer is fully transparent,
i. e. all the (very low) incident power
is transmitted through, while a giant R-wave 
of peak intensity $( u_{in} )_{pk}^2 \approx 2 \delta$
inside the layer is sustained by a tiny incident power,
$( u_{in} )_{min}^2$. 
If  $L \sqrt {2 \delta} / N \lambda > 1$,
the contrast ratio -- essentially a nonlinear resonator's
finesse, $Q$, is
%
\begin{equation}
Q = \frac{( u_{in} )_{pk}^2} {( u_{in} )_{min}^2 }
 \approx  
\frac{\exp ( 2 \pi L \sqrt {2 \delta} / N \lambda )} 
{ 128 \ \delta  } \gg 1
\tag{11}
\end{equation}
where $N$ is the number of R-waves in a layer;
the one with  $N = 1$ occurs after the first jump-up.
In the example for Fig. 1 ($\delta = 0.02$, $L = 10 \lambda$),
$ Q  \sim 10^7$.
In semi-infinite plasma, $Q$ is limited by absorption, see below.
It also decreases as $N$ increases;
the field profile for $N = 2$, 
is depicted in Fig. 1, point 3, curve 3 in the inset.

Only half of multi-steady-states are stable;
the stability condition
is that the EM-energy density
increases with the pumping,
i. e. $d W / d ( u_{in}^2 ) > 0$,
which also coincides with the condition
$d ( u_{ex}^2 ) / d ( u_{in}^2 ) > 0$, similar to [1].

One can view R-waves at a $N$-th stable branch
as a $N$-th order mode of a self-induced resonator,
with full transparency points marking the resonance.
The mirror-like sharp boundaries of plasma layer
enhance the resonances [10],
but do not constitute necessary condition.
Our numerical simulations using (4) 
showed that a layer with "soft" 
shoulders making $\sim$ 50\% of the entire layer length,
still exhibits a few hysteresises, and a
large number of self-induced resonances.

The fundamental manifestation of
the phenomenon transpires in 
a \emph{semi-infinite} plasma.
Only two kind of waves [1] in a lossless case
satisfy then the Sommerfeld condition -- no wave 
"comes back" from $\zeta \to \infty$ --
a traveling (7), $du / d \zeta \to 0$, 
and an evanescent (8) wave, $u \to 0$.
Our investigation of (2), to be 
published elsewhere, showed 
that the wave (7) is unstable
both in 2D\&3D-propagation
-- and, of (1) -- in temporal domain in 1D-case.
However, using (4), one can show
that even a \emph{steady} 1D-wave (7) is
unstable against small absorption
by replacing $\nu^2$ in $\epsilon_{pl}$ (3) 
with $\nu^2 ( 1  + i \alpha )$, 
where $\alpha =  ( \omega \tau )^{-1}$ is an absorption factor,
and $\tau$ is an electron momentum relaxation time
(typically $\alpha \ll 1$).
A condition for a hysteresis to emerge is then
$\alpha < \alpha_{cr} \approx e \delta$ [11], if $\delta \ll 1$.
Near $\alpha \sim \alpha_{cr}$, a jump-up
occurs at $ u_{in}^2 \approx \pi \delta$.
The ratio (11)
to sustain a single R-wave is limited now by 
$Q \approx  \delta / \alpha$, and can still be huge.
Hysteresises in reflection at
$\alpha = 10^{- 3}$ are shown in Fig. 2,
and the intensity profile for $u_{in}^2 > 2 \delta$ -- in the inset.
\begin{figure}
\includegraphics[angle=270,width=3in]{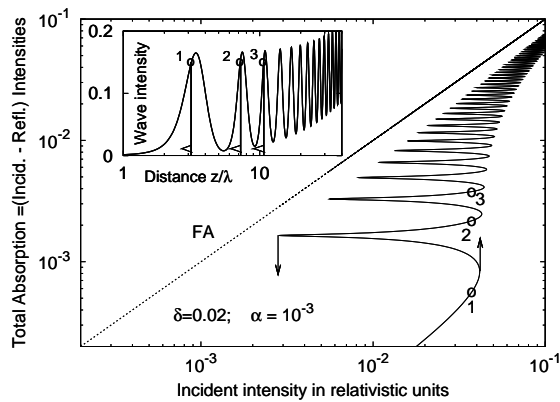}
\caption{Hysteretic absorption of light in a
semi-infinite plasma layer with absorption $\alpha$. 
$\textbf{FA}$ is a full absorption limit.
Points 1, 2, 3 mark a linear evanescent wave,
1-st, and 2-nd upper stable states respectively.
Arrows -- the same as in Fig. 1.
Inset: intensity profile; points and verticals 1-3
indicate locations of the plasma boundary 
for the respective points in the main plot;
$\triangleleft$'s show direction into plasma layer.
}
\end{figure}
An initially traveling wave develops oscillations
due to rising standing wave, which
eventually becomes a train of trapped R-waves, 
the last one being a quasi-soliton close to (8),
and then vanishes exponentially.
Reducing $\alpha$ pushes that 
last R-wave further back, but does not extinguish
retro-reflection from the R-wave train
at the crossover area deep inside plasma,
keeping the condition $u  \to 0$ at $\zeta \to \infty$.

Lab observation of the phenomena in plasma could be set up with 
e.\ g.\ jets of gas irradiated by a powerful $C O_2$ laser,
with a gas density controlled to reach a crossover point at
$\lambda_{ C O_2 } \approx 10 \mu m$;
the recent experiments [12] 
may actually indicate R-waves released by plasma expansion.
This process may be naturally
occurring in astrophysical environment in plasma sheets expelled
from a star (e. g. the Sun);
part of the star's radiation spectrum 
below the initial plasma frequency is 
powerful enough to penetrate into the layer
and be trapped as R-waves.
When the layer expands, they can 
be released as a burst of radiation.
It is also conceivable that the R-waves
trapping and consequent release may be part of the physics
of ball-lighting subjected to a powerful
radiation emitted by the main
lighting discharge in $mw$ and far infrared domains.
The R-waves might be used e.\ g.
for laser fusion to deposit laser power
much deeper into the fusion pallets;
or for heating the ionosphere layers by 
a powerful \emph{rf} radiation.

In conclusion, optical multi-hysteresises
may emerge in a plasma near critical
plasma frequency due to fundamental
relativistic mass-effect of electrons.
They may result in huge trapped,
or standing rogue waves with the intensity 
greatly exceeding that of pumping radiation.

This work is supported by the US AFOSR.



\begin{thebibliography} {90}
\bibitem{HYS}
This suggests an explanation of the hysteresis lacking 
in [2b]: an artificial nonlinearity used there was due to
small dielectric particles suspended in a liquid and had
large dissipation because of strong light scattering, $vs$ the 
experiments [2a] and [2c] with nearly transparent fluids.
\end{thebibliography}
\end{document}